\documentclass[10pt,conference]{IEEEtran}

\usepackage{cite}
\usepackage{amsmath,amssymb,amsfonts}
\usepackage{algorithmic}
\usepackage{graphicx}
\usepackage{textcomp}
\usepackage[usenames,dvipsnames]{xcolor}
\usepackage{url}
\usepackage[hidelinks,bookmarks=false]{hyperref}

\usepackage{listings}
\usepackage{fancyvrb}
\usepackage{booktabs}
\usepackage[absolute]{textpos}

\makeatletter
\def\endthebibliography{%
	\def\@noitemerr{\@latex@warning{Empty `thebibliography' environment}}%
	\endlist
}

\makeatother

\begin{document}

\title{Assessing the Impact of Refactoring Energy-Inefficient Code Patterns on Software Sustainability: An Industry Case Study\vspace{-0.4em}}

\author{
	\IEEEauthorblockN{Rohit Mehra$^\dagger$, Priyavanshi Pathania$^\dagger$, Vibhu Saujanya Sharma$^\dagger$, Vikrant Kaulgud$^\dagger$, Sanjay Podder$^\ddagger$, Adam P. Burden*}
	\IEEEauthorblockA{$^\dagger$Accenture Labs, India | $^\ddagger$Technology Sustainability Innovation, Accenture, India | *Accenture, USA\\
		\{rohit.a.mehra, priyavanshi.pathania, vibhu.sharma, vikrant.kaulgud, sanjay.podder, adam.p.burden\}@accenture.com\vspace{-0.4em}}
}

\maketitle

\begin{textblock*}{10cm}(1.5cm,26.2cm) 
	DOI: \url{https://doi.org/10.1109/ASE56229.2023.00205}
\end{textblock*}

\begin{abstract}

Advances in technologies like artificial intelligence and metaverse have led to a proliferation of software systems in business and everyday life. With this widespread penetration, the carbon emissions of software are rapidly growing as well, thereby negatively impacting the long-term sustainability of our environment. Hence, optimizing software from a sustainability standpoint becomes more crucial than ever. We believe that the adoption of automated tools that can identify energy-inefficient patterns in the code and guide appropriate refactoring can significantly assist in this optimization. In this extended abstract, we present an industry case study that evaluates the sustainability impact of refactoring energy-inefficient code patterns identified by automated software sustainability assessment tools for a large application. Preliminary results highlight a positive impact on the application's sustainability post-refactoring, leading to a 29\% decrease in per-user per-month energy consumption.

\end{abstract}


\section{Introduction}\label{introduction}

It is becoming increasingly evident that software and the corresponding software industry are significant and rapidly evolving contributors to global carbon emissions. Multiple studies have estimated that the internet and communications technology industry, which encompasses software and the corresponding hardware infrastructure, currently accounts for 2-7\% of global greenhouse gas emissions and is expected to increase to 14\% by 2040 \cite{unepccc, FREITAG2021100340}. The software code itself, and the best practices adopted during the development process, can have a significant impact on its carbon emissions. With the climate crisis looming, it is crucial to adopt corrective measures for reducing these rapidly growing emissions.

One of the primary reasons behind these high carbon emissions is the non-optimization of software from a sustainability perspective (specifically green/energy/emissions perspective) \cite{10.1145/3154384}. There are multiple inherent reasons like lack of awareness, hidden impact, etc. attributing to this non-optimization \cite{9793921, 10.1145/3350768.3350770}. However, we believe that the dearth of automated software sustainability assessment tools that can detect energy-inefficient patterns in code, gauge their sustainability impact, and offer potential remediation strategies, is one of the foremost \cite{7168335}. This becomes a major challenge for an environmentally-inclined practitioner who intends to refactor energy inefficiencies in her code but is (i) unaware of their existence (ii) unable to detect their precise location in a large industrial software, and (iii) lacks the required knowledge on how to correctly refactor them for making the software greener.

Despite their dearth, a few tools have recently tried to solve this problem. For example, CAST's Application Intelligence Platform (AIP) is a popular static code analyzer that analyzes the software for the presence of multiple types of inefficiencies/code-smells/violations, including performance, quality, reliability, security, etc., and across multiple technologies \cite{10.1145/1985793.1985893, castaip}. Recently, they have added support for detecting energy-inefficient patterns in code, referred to as \textit{Green IT Index}  \cite{Bessi:MeGSuS2016, castgreenitindex}. The \textit{Green IT Index} is a score ranging between 1-4 and is calculated based upon the number of detected energy-inefficient code patterns, where a score of 4.0 signifies the non-existence of any energy-inefficient patterns in code (a higher score corresponds to a greener software). These patterns are detected based upon a well-curated set of rules derived from open standards like CWE \cite{cwe}, ISO-5055 \cite{9734273}, etc. \cite{castrules}. These rules are aimed at detecting patterns that drive excessive use of hardware resources, thereby contributing to excessive energy consumption. Moreover, post-detection, CAST AIP also recommends corrective measures for refactoring the detected energy-inefficient patterns. Finally, in contrast to CAST AIP, a few other recent tools that do exist in this domain, for example - ecoCode \cite{10.1145/3551349.3559518}, are still in the early stages of development, limited to specific technologies, and further limited to a minuscule catalog of energy inefficiencies that they can detect, which currently renders them ineffective for industry adoption.

Despite their claims of making the software greener, there is a lack of reported case studies to validate these claims. For instance, what if the utilization of such tools leads to an ``increase'' in energy consumption for a specific industry application? Conducting case studies in these scenarios can offer preliminary insights into the efficacy of these tools and influence their adoption or non-adoption from a software sustainability perspective.

In this paper, we study the sustainability impact of refactoring energy-inefficient code patterns identified by software sustainability assessment tools. For this study, we processed a large internal application using CAST AIP and implemented the suggested refactorings with our development team. Initial results demonstrate a positive impact on the application's sustainability, resulting in lower energy consumption and reduced carbon emissions.
\section{Impact Of Refactoring Energy-Inefficient Code Patterns On Software Sustainability}\label{experiments}

\begin{figure}[t]
	\centering
	\includegraphics[width=1.0\linewidth]{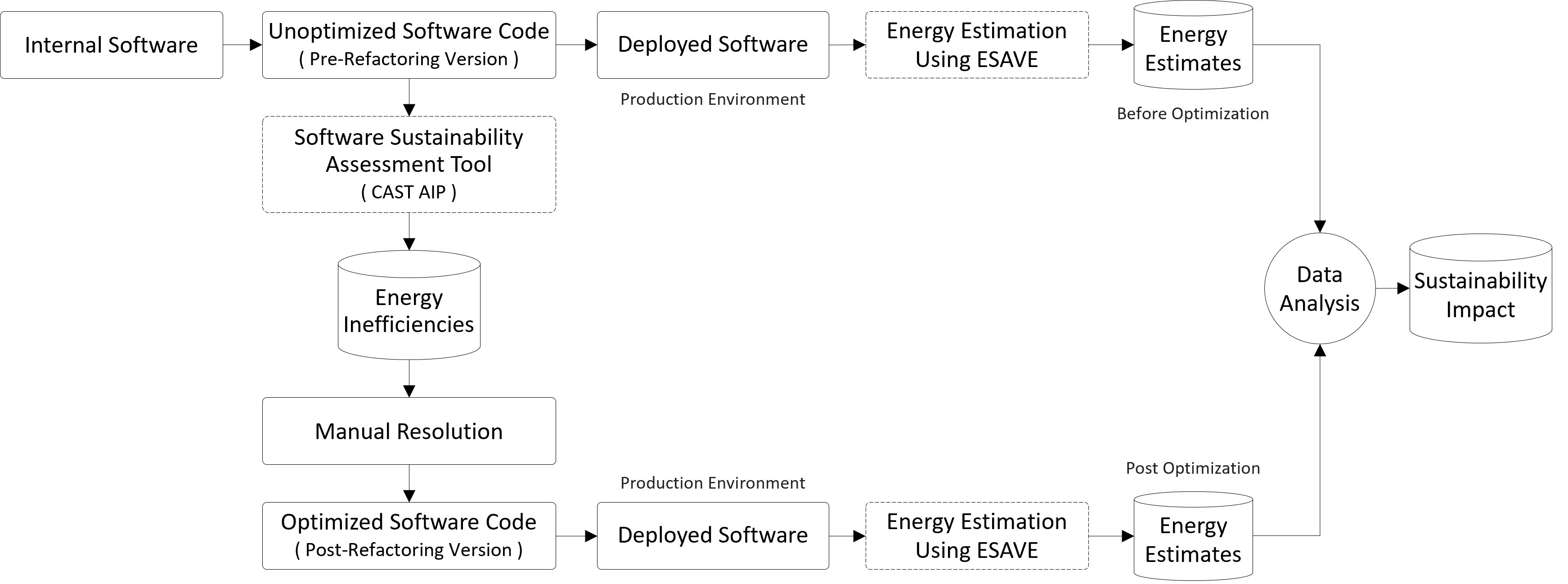}
	\caption{Schematic diagram for the study methodology.}
	\label{fig:methodology}
\end{figure}

For the study, we selected a large internal application that has been developed in-house. It serves as a digital solution planning platform with multiple accessible benchmarks that enables solution architects to swiftly create competitive client solutions. The application is used by over 300 solution architects every month. Since it was in the maintenance phase, the application proved to be an ideal candidate for this study, as no new functionalities or bug fixes were being actively pursued that could potentially tamper with the eventual results.

\subsection{Study Methodology}

Figure \ref{fig:methodology} presents the methodology of our study. To identify the energy inefficiencies in the application, we leveraged the version of the application that existed as of January 2022 and analyzed its source code using the CAST AIP tool. This version will be referred to as the \textit{pre-refactoring} version. Tool analysis revealed 15130 energy inefficiencies in the \textit{pre-refactoring} version, with a Green IT Index of 2.81. These energy inefficiencies span multiple technical categories including programming practices, secure coding, etc., and were further marked as critical/non-critical. These inefficiencies were thereafter refactored by our development team for thirty days, by leveraging the guidance and example code snippets provided by the tool. The eventual version will be referred to as the \textit{post-refactoring} version. The \textit{post-refactoring} version was thereafter analyzed by the tool to gauge the updated energy inefficiencies and Green IT Index. A few key metrics of both versions are highlighted in Table \ref{tab:table_results}.

For computing the sustainability metrics (energy consumption and carbon emissions), we leveraged our prior research on energy estimation of servers and virtual machines called ESAVE \cite{esave}. The energy numbers estimated by ESAVE were then converted into equivalent carbon emissions using average monthly regional grid intensities provided by Electricity Maps \cite{electricitymaps}. To estimate the sustainability metrics, both the versions were deployed on the same production environment, comprising predominantly of five virtual machines on Microsoft Azure cloud \cite{Copeland2015}, for an entire month (January 2022 and March 2022, respectively), and the telemetry required for energy estimation was manually extracted. For example, underlying CPU configuration, hourly CPU utilization, deployment region, etc.

\subsection{Initial Results}

Table 1 summarizes the findings from the study. According to results, the \textit{post-refactoring} version demonstrated 60.82\% fewer energy inefficiencies as compared to the \textit{pre-refactoring} version, leading to a 25.62\% increase in the Green IT Index. Further analysis reveals that this improvement led to a 3.26\% decrease in the absolute monthly energy consumption. While this may seem insignificant, we further observed a 28.75\% decrease in the per-user per-month energy consumption. The latter was calculated to normalize the energy consumption numbers since the number of users was 35.71\% higher for the post-refactoring version, which might have an impact on the absolute energy consumption. This normalization is also in line with the ``software carbon intensity'' metric, proposed by the Green Software Foundation \cite{sci}. Finally, we also observed a 40.23\% decrease in carbon emissions, which is higher than the energy reduction percentage. This is predominantly due to the change in the average regional grid intensity between the two observation periods, while the deployment region remained the same. These results highlight a positive impact of refactoring energy-inefficient code patterns, as identified by the automated software sustainability assessment tool, on software sustainability for a large industry application.

\begin{table}[t]
	\ttfamily
	\centering
	\caption{Comparing application metrics, pre and post refactoring.}
	\label{tab:table_results}
	\vspace{0.0em}
	{\small
		\resizebox{\columnwidth}{!}{%
		\begin{tabular}{@{}lrrr@{}}
		\toprule
												 & \textbf{Pre Refactoring} & \textbf{Post Refactoring} & \multicolumn{1}{l}{}   \\
												 & \textbf{January 2022}    & \textbf{March 2022}       & \textbf{\% Difference} \\ \midrule
		\textbf{General Metrics}                                                                                                 \\
		End Users                                & 336                      & 456                       & \textbf{+ 35.71 \%}    \\
		Lines of Code                            & 1.98 M                   & 2.07 M                    & \textbf{+ 04.54 \%}    \\ \midrule
		\textbf{CAST AIP Metrics}                                                                                                \\
		Detected Energy Inefficiencies           & 15130                    & 5928                      & \textbf{- 60.82 \%}    \\
		Green IT Index                           & 2.81                     & 3.53                      & \textbf{+ 25.62 \%}    \\ \midrule
		\textbf{Sustainability Metrics}                                                                                          \\
		Energy Consumption (Monthly)             & 53.74 kWh                & 51.99 kWh                 & \textbf{- 03.26 \%}    \\
		Energy Consumption (Per User Per Month)  & 0.160 kWh                & 0.114 kWh                 & \textbf{- 28.75 \%}    \\
		Carbon Emissions (Monthly)               & 25.74 KgCO2e             & 21.11 KgCO2e              & \textbf{- 17.99 \%}    \\
		Carbon Emissions (Per User Per Month)    & 0.077 KgCO2e             & 0.046 KgCO2e              & \textbf{- 40.23 \%}    \\ \bottomrule
		\end{tabular}
		}
	}
	\vspace{0.0em}
\end{table}
\section{Conclusion and Future Directions}\label{conclusion}

In this paper, we showcased an industry case study of applying software sustainability assessment tools to a large application. The case study demonstrates a positive impact on key sustainability metrics following the adoption of energy-inefficient code refactoring suggestions provided by the tool. In the future, we plan to expand this study to encompass additional applications and explore other emerging tools in this field. Furthermore, we acknowledge the need for further in-depth explorations, as similar results may not apply universally to all tools and types of energy-inefficient code patterns. We firmly believe that these types of tools can play a significant role in managing the increasingly significant carbon footprint of software systems and guide practitioners toward energy-efficient coding practices, thereby enabling the true notion of Green DevOps. Additionally, we strongly encourage the software engineering community to embrace these tools, fostering focused and collaborative efforts toward achieving a sustainable future.

\section*{Acknowledgement}

The authors would like to thank their collaborators in Advance Technology Centers, India for their generous support throughout the journey of this case study.

\bibliographystyle{IEEEtran}
\bibliography{Bibliography} 

\end{document}